\documentclass[num-refs]{wiley-article}
\usepackage{color}
\usepackage[normalem]{ulem}  
  
\usepackage{amsfonts} 

\usepackage{multirow}
\usepackage[scr=esstix,cal=boondox]{mathalfa} 
\usepackage{hyperref}
\hypersetup{
    colorlinks=true,
    urlcolor=black,
    citecolor=blue,
}
\usepackage[ruled,vlined]{algorithm2e}

\renewcommand{\v}[1]{\ensuremath{\mathbf{#1}}}
\newcommand{\gv}[1]{\bm{#1}}
\def\[#1\]{\begin{equation}#1\end{equation}} 

\usepackage[nice]{nicefrac}

\papertype{Research Article}

\begin{document}
\runningauthor{Sacco M. A. et al. }

\title{Online machine-learning forecast uncertainty estimation for sequential data assimilation}
\author[1,2]{Maximiliano A. Sacco}
\author[3,4]{Manuel Pulido}
\author[2,3,5]{Juan J. Ruiz}
\author[6,7]{Pierre Tandeo}

\affil[1]{Servicio Meteorol\'ogico Nacional, Buenos Aires, Argentina}
\affil[2]{Departamento de Ciencias de la Atm\'osfera y los Oc\'eanos, Facultad de Ciencias Exactas y Naturales, Universidad de Buenos Aires, Buenos Aires, Argentina}
\affil[3]{CNRS-IRD-CONICET-UBA, Instituto Franco-Argentino para el Estudio del Clima y sus Impactos (IRL 3351 IFAECI), Buenos Aires, Argentina}
\affil[4]{Departamento de F\'isica - Facultad Ciencias Exactas y Naturales y Agrimensura, Universidad Nacional del Nordeste, Corrientes, Argentina}
\affil[5]{Centro de Investigaciones del Mar y la Atm\'osfera, Facultad de Ciencias Exactas y Naturales, Universidad de Buenos Aires, CONICET-UBA, Buenos Aires, Argentina}
\affil[6]{IMT Atlantique, Lab-STICC, UMR CNRS 6285, 29238, France}
\affil[7]{Odyssey, Inria/IMT, France}

\corraddress{msacco@smn.gob.ar}
\corremail{msacco@smn.gob.ar}
\maketitle

\begin{abstract}
Quantifying forecast uncertainty is a key aspect of state-of-the-art numerical weather prediction and data assimilation systems. Ensemble-based data assimilation systems incorporate state-dependent uncertainty quantification based on multiple model integrations. However, this approach is demanding in terms of computations and development. In this work a machine learning method is presented based on convolutional neural networks that estimates the state-dependent forecast uncertainty represented by the forecast error covariance matrix using a single dynamical model integration. This is achieved by the use of a loss function that takes into account the fact that the forecast errors are heterodastic. The performance of this approach is examined within a hybrid data assimilation method  that combines a Kalman-like analysis update and the machine learning based estimation of a state-dependent forecast error covariance matrix. 
Observing system simulation experiments are conducted using the Lorenz'96 model as a proof-of-concept. The promising results show that the machine learning method is able to predict precise values of the forecast covariance matrix in relatively high-dimensional states. Moreover, the hybrid data assimilation method shows similar performance to the ensemble Kalman filter outperforming it when the ensembles are relatively small. 


\end{abstract}

\section{Introduction}

Quantifying forecast uncertainty is a key aspect of data assimilation (DA) systems. In particular most DA methods rely on an accurate estimation of the forecast mean and  error covariance matrix. Together they describe the  probability density function under the assumption that errors are unbiased and Gaussian.


Data assimilation approaches such as optimal interpolation (OI, \citealt{gandin1965}) or 3-dimensional variational methods (3DVar, \citealt{parrishandderber92}) assumes that the forecast error covariance matrix is independent of the state of the system. Currently, DA methods that provide an implicit (e.g. 4-dimensional varitional methods, 4DVAR, \citealt{rabier2000}) or explicit (e.g. ensemble Kalman filters, EnKFs, \citealt{houtekamerzhang2016}, particle filters, PFs, \citealt{vanleeuwen2019}) or hybrid \citep{bannister2017}, estimation of the state dependent forecast probability density function produce a remarkable improvement in the accuracy of the initial conditions and of the forecast skill \citep{kalnay2003,carrassi2018}. However, these improvements come at the expense of a significant increase in the computational cost. Moreover, even when state-dependent error covariances are well represented, an accurate estimation of the contribution of model errors to the forecast error covariance in 4Dvar and EnKF frameworks is still challenging \citep{tandeoetal2020}. 





Recently, machine learning methods---trainable statistical models that can represent complex functional dependencies among different groups of variables given a large enough dataset--- have emerged as a promising alternative to estimate the forecast uncertainty (e.g. \citealt{tandeo2015combining,oualaetal2018,wang,camporeale2,gronquist2019,irrgangetal2020,grooms2021,Sacco2022}, among others). These methods provide an accurate estimation of the forecast uncertainty at a relatively low computational cost (i.e., without the need of multiple integrations of the numerical model or its adjoint). Moreover, apart from the forecast uncertainty quantification, some of these methods capture also an estimation of the uncertainty associated with model errors, which are difficult to estimate (e.g., \citealt{camporeale1,oualaetal2018,wang,Sacco2022}). These methods rely on uncertainty-aware loss functions allowing the ML algorithms to learn the error statistics directly from the data (see for example, \citealt[Chapter~5.6]{bishop2006}). 

Most of these works have focused on the estimation of the forecast error variance (e.g. \citealt{wang,camporeale1,gronquist2019,irrgangetal2020,Sacco2022} among others). However, the estimation of the full error covariance structure is essential for data assimilation. \cite{grooms2021} estimated the full covariance structure based on a machine learning method designed to provide an ensemble of perturbations of the state variables that represents possible realizations of the forecast error. This approach emulates the one used in ensemble forecasting but without the need to integrate the computationally demanding numerical model to generate the ensemble members. \cite{lguensat2017} replace the numerical model for a surrogate model based on a local linear analog regression, thus significantly reducing the computational cost associated with the numerical integration of the ensemble. \cite{oualaetal2018} use a neural network and a Gaussian likelihood based loss function to estimate a diagonal error covariance in a subspace defined by the leading principal components of the state variables resulting in an approximation of the full forecast error covariance. 

The estimation of a full error covariance matrix from data has been investigated in other contexts. \cite{williams96} used a neural network to estimate the parameters of a multivariate Gaussian distribution. \cite{Humphrey2015} presented a 
parametric covariance prediction for heteroscedastic noise and \cite{Liuetal2018} implemented a deep learning model for the inference of the observation error covariance matrix and applied it to position estimation for navigation applications. In these cases, a Cholesky decomposition of the covariance matrix is estimated based on the Gaussian likelihood.


The use of machine learning (ML) techniques in the context of data assimilation have been discussed in several works. The similarities between DA and ML and their potential synergism has been introduced in \cite{hsiehandtang1998} and reviewed in \cite{cheng2023machine}. \cite{bocquetetal2019,brajard2020,farchietal2021,farchietal2022} proposed a framework in which machine learning is used for the estimation of the system dynamics and to represent model errors, while data assimilation provides an online continuous optimization of the data-driven model. Along the same line, \cite{brajard2021, farchietal2021b}, use a data assimilation approach to train an ML-based parameterization of the effect of unresolved scale dynamics within a numerical model. 
Other approaches aimed to a replacement of the full DA system by a neural network as in \cite{Harteretal2008}. In this approach the authors train a neural network that learns, from a given DA system, the magnitude and spatial patterns of the state update introduced by the observations. \cite{buizzaetal2021} introduced the name "data learning" to describe several examples in which ML and DA can be combined to overcome their mutual weaknesses. 

Relatively few works investigated how ML-based forecast error covariance estimation, can be coupled with a DA system (e.g., \citealt{lguensat2017,oualaetal2018}). In particular, \cite{oualaetal2018}, coupled a neural network-based estimation of the forecast error covariance with a Kalman like analysis update in a high dimensional state space and compared the results with an ensemble Kalman filter approach and the analog-based approach of \cite{lguensat2017} with promising results. In this work. we  investigate the impact on the quality of the estimation of the state of a dynamical system, particularly when a localized version of the full error covariance matrix is directly estimated using an extension of the novel loss function presented in \cite{Sacco2022}.  The neural network covariance estimation uses a single forecast as input variable which is obtained by means of a numerical model.  The stability of the method is investigated by performing several assimilation cycles. This \textbf(U)ncertainty estimation with \textbf{n}eural \textbf{n}etworks is integrated with a Kalman filter-based data assimilation system, forming a hybrid technique referred as UnnKF.  

This work is structured as follows: Section \ref{SEC:METH} describes the different approaches for the estimation of the forecast error covariance including a brief review of the ensemble Kalman filter and the experimental settings. The design of all the experiments that were carried out in this work are described in Section \ref{SEC:EXP}. Section \ref{SEC:RES} analyzes the results obtained. Section \ref{SEC:CONC} draws the main conclusions of this work as well as a discussion of future perspectives.


\section{Methodology}
\label{SEC:METH}

\subsection{Sequential data assimilation}
\label{SEC:OI}

In a sequential data assimilation cycle, we aim to estimate the state of a dynamical system at regular time intervals, by combining the information provided by a surrogate numerical model and a set of partial and noisy observations \citep{carrassi2018}.
We start by considering a chaotic dynamical system, represented via the following Markov process
\begin{equation}
\label{EQU:sdm1}
\v x_{k} = \mathcal{M}_{k:k-1}(\v x_{k-1})+\gv \eta_k,
\end{equation}
where $\v x_{k}$ is an $N_x$-dimensional vector representing the state of the system at time $k$, $\mathcal{M}_{k:k-1}$ is a known nonlinear and chaotic imperfect model of the system dynamics that maps the state at time $k-1$ into time $k$, and $\gv\eta_k$ represents the discrepancy between $x_{k}$ and $\mathcal{M}_{k:k-1}(\v x_{k-1})$ due to the model imperfection (i.e., the model error). In this work, we assume that the model error is a random variable sampled from a Gaussian probability distribution. 

Given a pointwise estimation of the state of the system at time $k-1$ ($\v x^a_{k-1}$) a deterministic forecast of the state at  time $k$ can be obtained by integrating the dynamical model and neglecting model errors,

\begin{equation}
\label{EQU:detfor}
\v x^{f}_{k} =  
\mathcal{M}_{k:k-1}(\v x^{a}_{k-1}),
\end{equation}
Forecasts for longer lead times can be obtained by a recursive application of the numerical model. 
The forecast error can be defined as:  
\begin{equation}
\label{EQU:forerr}
\gv \epsilon_{k}^{f} = \v x^{f}_{k} - \v x^{t}_{k},
\end{equation}
where $\v x^{t}_{k}$ is the unknown true state of the system at time $k$. Forecast errors are the consequence of an imperfect estimation of the state of the system at time $k-1$ and model errors. The magnitude and structure of both contributions to the forecast error depend strongly on the state, so that the structure and magnitude of the component of the forecast error covariance matrix at time $k$ ($\gv P^{f}_{k} = [\gv\epsilon_{k}^{f} {\gv\epsilon_{k}^{f}}^\top]$) are a function of the state. Data assimilation methods rely on the assumption that these errors have zero-mean, which is not usually the case. 

The state of the system is related to the observable quantities through the  observation equation, 
\begin{equation}
\label{EQU:obsmodel}
\v{y}_{k} = \mathcal{H}( \v x^{t}_{k} ) + \gv \nu_{k},
\end{equation}
where $\v y_{k}$ is the $N_y$-dimensional vector containing the observable quantities, $\mathcal{H}$ is the observation operator (i.e. the function mapping state variables into the observation space) and $\gv \nu_k$ is the observation error which is assumed to be drawn from a Gaussian distribution with zero-mean and known covariance denoted $\v R_k$.

Given the forecast ($\v x^{f}_{k}$), a set of observations ($\v y_{k}$), and assuming that their errors are unbiased , the best linear estimator that minimizes the root mean square error with respect to the true state of the system is given by:

\begin{subequations}

\begin{align}
&\v{x}^a_{k} = \v{x}^f_{k} + \gv{K}(\v{y}_k - \mathcal{H}(\v{x}^f_{k}) ), \label{EQU:analysis_a}\\
&\gv{K} = \gv{P}^f_k \gv{H}^\top(\gv{H} \gv{P}^f_k \gv{H}^\top + \gv{R}_k)^{-1},\label{EQU:analysis_b} 
\end{align}
\label{EQU:analysis}
\end{subequations}
where $\v{x}^a_{k}$ is the estimation of the system state (a.k.a the analysis) at time $k$, $\gv{H}$ is the tangent linear approximation of the observation operator and $\gv K$ is the Kalman gain matrix which projects and weights the discrepancy between the observations and the forecasted observed quantities into the state space. This estimate of the state is also the maximum likelihood estimation of the state of the system under the assumption that the PDFs of the forecast errors and observation errors are both zero mean and Gaussian \citep{carrassi2018}. Depending on the forecast covariance, $\gv{P}^f_k$, Eq. \ref{EQU:analysis} may represent an optimal interpolation or an extended Kalman filter. In the optimal interpolation approach, $\gv{P}^f_k$ is usually assumed to be known \textit{a priori} and state-independent, while in the extended Kalman filter, the time evolution of $\gv{P}^f_k$ is computed using the tangent linear approximation of the numerical model. 

Once an estimation of the system state is obtained at time $k$, the numerical model (Eq. \ref{EQU:detfor}) can be used to forecast the state of the system for the next time, and the cycle can be repeated every time a new set of observations becomes available. The accuracy of the state estimation depends strongly on the accuracy of the error covariance matrices $\gv P^f_k$ and $\gv R_k$ whose estimation is arguably one of the most challenging aspects of DA systems \citep{tandeoetal2020}.

\subsection{The ensemble Kalman filter}
\label{SEC:ENKF}

The ensemble Kalman filter (EnKF) is one of the most broadly used methods to incorporate the state-dependence of the forecast error covariance matrix in data assimilation applications. In this work, the EnKF is used to generate the database for the training of the machine learning method and is used as a benchmark for the evaluation of the proposed machine learning-based algorithms. For completeness, we  briefly describe this technique here. 

If we have a sample of states drawn from the probability distribution of the analysis state at time $k-1$ $( \v x^{a,(n)}_{k-1} )$, for $n \in 1 ... N_e$ with $N_e$ the ensemble size, the sample covariance of the forecast at time $k$ can be estimated by evolving the individual ensemble members from time $k-1$ to time $k$ through the non-linear model equations:
\begin{align}
  \label{EQU:MODEL1}
  \v x^{f,(n)}_{k} =  \mathcal{M}^{(n)}_{k:k-1}\left(\v x^{a,(n)}_{k-1}\right) + \hat{\gv\eta}^{(n)}_{k},
\end{align}
where $\v x^{f,(n)}_{k}$ are the evolved ensemble members. 
The forecast ensemble mean at time $k$, $\overline{\v x}^{f}_{k} = \frac{1}{N_e} \sum_{n=1}^{N_e} \v x^{f,(n)}_{k}$ provides a pointwise estimation of the state. 
Along this line, the forecast error covariance can be estimated from the forecast state sample, 

\begin{equation}
\label{EQU:SIGMA}
\hat{\gv P}^{f}_{k} = \frac{1}{(N_e-1)}\sum_{n=1}^{N_e} \left(\v x^{f,(n)}_{k}-\overline{\v x}^{f}_{k}\right)\left(\v x^{f,(n)}_{k}-\overline{\v x}^{f}_{k}\right)^\top .
\end{equation}

In the stochastic implementation of the EnKF \citep{burgers1998}, the ensemble members are updated using Equation \ref{EQU:analysis_a}, in which $\v y_k$ is replaced by $\v y^{(n)}_k=\v y_k + \v \nu^{(n)}_k$, with $\nu^{(n)}_k \sim \mathcal{N}(\v 0,\gv R_k)$ and with $\v P^f_k$ given by Eq. \ref{EQU:SIGMA}.

In physical systems, the covariance between variables corresponding to locations that are far away in physical space are close to zero. In the EnKF, due to the presence of sampling errors, covariances between distant variables can be significantly different from 0, particularly when a small ensemble is used. In this case, a covariance localization approach can be used to damp the magnitude of the spurious covariances. These methods usually multiply the estimated covariances by a factor that decreases with the physical distance between the two variables \citep{hamilletal2001}.

In this work, the stochastic EnKF was chosen over deterministic filters such as the LETKF \citep{HUNT07} since in these filters ensemble members are not equi-probable since some members are persistently associated with larger departures from the ensemble mean. This effect has already been reported by \cite{amezcua2012} and  found in a realistic experiment by \cite{kondoandmiyoshi2019}. This  affects negatively the training of the neural network models used in this work. The stochastic EnKF, because of the random sampling in the update of each ensemble member, does not suffer from this problem. Also, we note that the fine-tuned localized stochastic EnKF and the LETKF had the same performance in terms of RMSE in the conducted experiments.


\subsection{Uncertainty estimate with neural network for data assimilation}
\label{SEC:COUPLING}

The likelihood function of the Gaussian distribution may be used as a loss function to train a neural network to learn the state-dependent covariance matrix.  However, estimating a full error covariance matrix is difficult and computationally expensive to train due to the covariance matrix inversion in the evaluation of the likelihood function. The use of the Cholesky decomposition of the covariance matrix or its inverse, to ensure that the obtained matrix is positive semidefinite, have been proposed (e.g. \citealt{williams96,Liuetal2018,Humphrey2015}) along with the definition of the cost function in terms of the precision matrix to avoid performing the inversion of the covariance matrix in its computation. However, in preliminary experiments, the covariance estimated in this way suffers from serious numerical instability problems when coupled with a data assimilation cycle with state space dimensions in the order of $10^2$. An alternative solution was proposed by \cite{oualaetal2018} who assumes the covariance matrix to be diagonal in the space defined by the leading principal components of the state variables. In this space the problem reduces to the estimation of the variance while a full covariance matrix can be obtained in the original state space. 

\subsubsection{Extended-MSE loss-function for covariance estimation} 

The loss function we use was originally presented in \cite{Sacco2022} for variance estimation. The name extended-MSE or simply eMSE was originally proposed because this technique uses the mean squared error equation for training, but instead of using the training target directly, it uses an on-line estimate of the forecast error. In this work, we extend the use of this loss function for a full covariance estimation and we use it into a DA framework.

The estimation of the forecast error requires an approximation of the true state (Eq. \ref{EQU:forerr}),  
\begin{equation}
\label{EQU:eMSE_error}
\gv \epsilon_k^f \approx  \textbf{x}^f_k -  \hat{\textbf{x}}^t_k.
\end{equation}

The approximation of the true state $\hat{\textbf{x}}^t_k$ could be taken to be, for instance, the mean analysis provided that the analysis error is much smaller than the forecast error (i.e., the analysis is closer to the true state than the forecast). We note that under this approximation the trace of the analysis covariance is assumed to be significantly smaller than the forecast covariance. 
Further choices of proxies for model forecast error are discussed in Section \ref{SEC:PROXY} and will be evaluated in the experiments.

This forecast error can be used to generate a state-dependent training matrix as 
\begin{equation}
\label{EQU:eMSE_cov}
\gv\epsilon^f_k (\gv\epsilon_k^f)^\top =(\textbf{x}^f_k - \hat{\textbf{x}}^t_k)(\textbf{x}^f_k - \hat{\textbf{x}}^t_k)^\top.
\end{equation}

The predicted covariance by the neural network is represented by
\[\tilde{\gv \Sigma}_k=\mathcal F_{NN}(\textbf{x}^f_k, \textbf{x}^a_{k-1};\gv\theta),\]
where $\mathcal F_{NN}$ is the neural network, $\gv\theta$ its parameters and $\{\textbf{x}^f_k, \textbf{x}^a_{k-1} \} $ its input data. In \cite{Sacco2022}, it was shown that using $\{\textbf{x}^f_k, \textbf{x}^a_{k-1} \} $ as inputs to the network improved the estimation of the mean and the variance of the state variables with respect to $\{\textbf{x}^f_k\}$. Similar results were obtained for the estimation of the full forecast covariance matrix $\tilde{\gv \Sigma}_k$ (not shown). 

\begin{figure}[hbt!]
\centering
        \includegraphics[width=.5\textwidth]{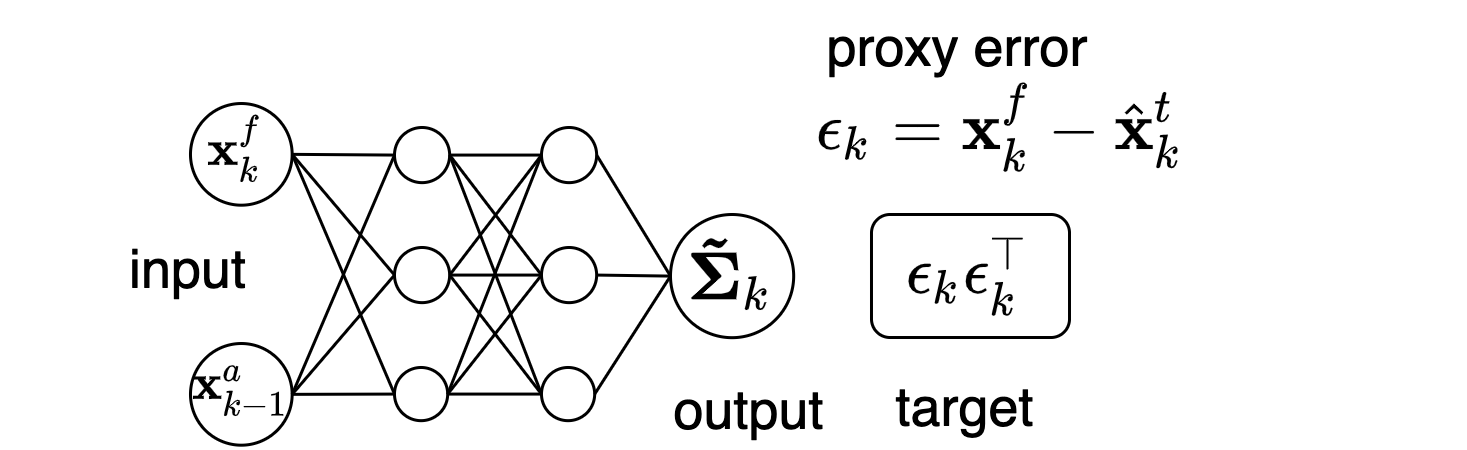}
        \caption{ ANN training scheme. The training error is determined by the Frobenius norm between  $\tilde{\gv \Sigma}_k$ and  the training matrix $\gv\epsilon^f_k (\gv\epsilon_k^f)^\top$ which is estimated from the approximated forecast error. }
\label{FIG:TRAINNET}
\end{figure}

Then, the loss function used for training (schematized in Figure \ref{FIG:TRAINNET}) is the square of Frobenius norm between the neural network output $\tilde{\gv \Sigma}_k$, and the training target $\gv\epsilon^f_k (\gv\epsilon_k^f)^\top$,
\begin{equation}
\label{EQU:eMSE_MSE}
\mathcal{L}(\gv \epsilon_k^f,\tilde{\gv \Sigma}_k)= \Big\lVert  \tilde{\gv \Sigma}_k -  \gv\epsilon^f_k (\gv\epsilon_k^f)^\top \Big\rVert_F = \sum^{N_x,N_x}_{i,j=0} (\tilde{\gv \Sigma}^{(i,j)}_k-{[\gv\epsilon^f_k (\gv\epsilon_k^f)^\top]}^{(i,j)})^2.
\end{equation}

In the EnKF method, localization methods in the covariance matrix are used to alleviate  sampling error. The same idea can be used to filter out spurious covariances between distant variables in the estimated covariance by applying a localization matrix $\gv C$ to the training target to force the decay of the estimated covariances with increasing distance in the physical space. As in EnKF, the structure of the localization matrix is a design decision based on knowledge of the dynamics of the problem or on empirical results. Based on this idea, the loss function is modified to

\begin{equation}
\label{EQU:LeMSE_MSE}
\mathcal{L}(\gv {\epsilon}^f_k,\tilde{\gv \Sigma}_k)= \lVert  \tilde{\gv \Sigma}_k - \gv C \circ [ \gv\epsilon^f_k (\gv\epsilon_k^f)^\top] \rVert_F 
\end{equation}
 where matrix $\gv C$ is assumed to be known a priori from the knowledgement of the dynamical interactions and selects the elements of the covariance matrix that will be estimated by the network, and $\circ$ is the element-wise product. Consistently,  all the elements of the output matrix ($\tilde{\gv \Sigma}_k$) corresponding to 0 values in $\gv C$ are removed by varying the size of the output layer of the network (see Sec. \ref{SEC:ARCH}). In this way, we reduce the number of training parameters and limit the computation of the covariances to only the selected subdiagonals (i.e. $\tilde{\gv \Sigma}_k$ may be represented by a band matrix).

\subsubsection{Data assimilation process}

\begin{figure}[hbt!]
\centering
        \includegraphics[width=.5\textwidth]{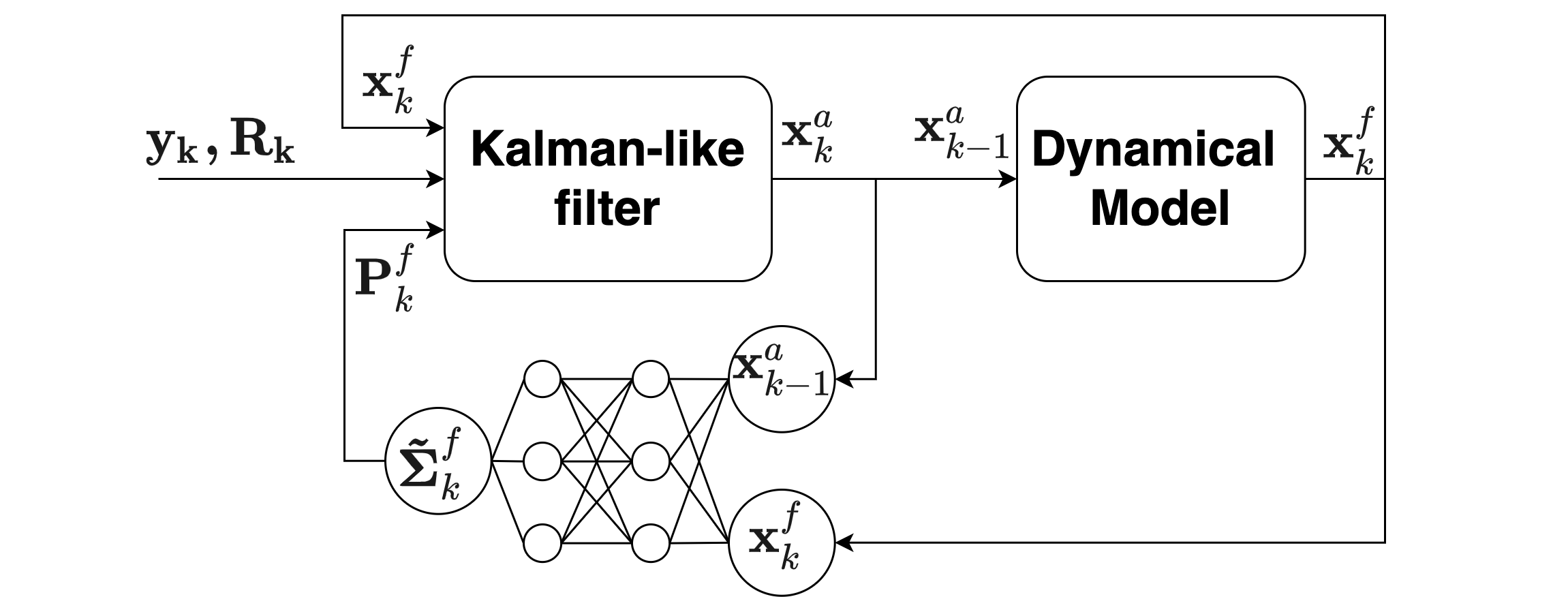}
        \caption{Schematic representation of an UnnKF assimilation cycle (see the text for details).  }
\label{FIG:ASSIM}
\end{figure}

Figure \ref{FIG:ASSIM} shows a schematic representation of the hybrid data assimilation cycle. At each assimilation cycle, the numerical model is initialized with the analysis of the previous cycle ($\textbf{x}^a_{k-1}$) providing  a deterministic forecast state, $\textbf{x}^f_k=\mathcal M(\textbf{x}^a_{k-1})$. The forecast and its corresponding analysis are used as inputs to the neural network to obtain an estimation of the forecast error covariance $\textbf{P}^f_k\approx\tilde{\gv \Sigma}_k=\mathcal F_{NN}(\textbf{x}^a_{k-1},\textbf{x}^f_k;\theta)$ which we plug into Eq. \ref{EQU:analysis_a} to obtain the analysis at time $k$, $\textbf{x}^a_{k}$. This in turn is used as initial condition to produce the forecast for the next assimilation cycle. 

This approach uses a single forecast from a numerical dynamical model to propagate the information on the state of the system from time $k-1$ to time $k$ (as in optimal interpolation or 3-dimensional variational approaches), but it uses a time-dependent estimation of the forecast error covariance matrix as in the EnKF. However, instead of using an ensemble of forecasts, 
the full covariance matrix is estimated by a neural network. 


The analysis update given by Eq. \ref{EQU:analysis_a}  is quite sensitive to the quality of the estimated forecast error covariance. For instance, if the diagonal terms are overestimated, the analysis tends to overfit the observations. In addition, if the subdiagonal elements are not well estimated, the information of the observations does not propagate properly to the unobserved variables of the system. In the optimal interpolation or 3-dimensional variational approaches, it is assumed that the covariance of the forecast does not depend on the state of the system. In the EnKF, the state dependence is taken into account, but sampling errors due to a limited ensemble size can affect its accuracy \citep{hamilletal2001}. In the case of UnnKF, the quality of the covariance will be determined by the ability of the neural network to learn the relationship between the state of the system and the associated uncertainty, in our case, the covariance.

In \cite{Sacco2022} the estimation of the forecast error variance was done in combination with an estimation of the state-dependent forecast bias. In this work, our main goal is to evaluate the accuracy and effectiveness of the covariance estimation in the context of a sequential data assimilation system. Although the forecast bias correction could improve the performance of the assimilation, we do not include it as part of the experiments in this work, since it could hide the sensitivity of the analysis error to the accuracy of the covariance estimation. In other words, all the improvements with respect to a fixed covariance optimal interpolation in this work  can be ascribed to the neural network covariance estimation.



\section{Experimental design}
\label{SEC:EXP}

\subsection{Dataset generation}
\label{SEC:L96}

For the generation of the datasets used to train the neural networks and to validate their performance we used a simplified data assimilation system based on the Lorenz'96 \citep{lorenz96} dynamical model. This is a simple chaotic model widely used in proof-of-concept experiments in the data assimilation community (e.g. \citealt{Stanley2021,brajard2020,lguensat2017,Koji2014}).

In particular, the two-scale Lorenz model \citep{lorenz96} is used to represent the evolution of the unknown nature state. This two-scale system allows us to represent the essence of multiple spatio-temporal scale systems such as the atmosphere or the ocean. The large and small-scale dynamical variables are governed by 
\begin{equation}
\begin{split}
  \frac{dx_{(i)}}{dt}= &-x_{(i-1)}(x_{(i-2)}-x_{(i-1)})-x_{(i)}+F-\frac{hc}{b}\sum_{j=J(i-1)+1}^{iJ} y_{(j)}\\
  \frac{dy_{(j)}}{dt}= & -cb \,y_{(j+1)}(y_{(j+2)}-y_{(j-1)})-c \,y_{(j)}+\frac{hc}{b}x_{(\mathrm{int}[(j-1)/J]+1)},
  \end{split}
\label{Lorenz2S}
\end{equation}
where $x_{(i)}$ is the $i-th$ component of the slow dynamics state vector $\v x$, and $y_{(j)}$ is the $j-th$ component of the fast-dynamics state vector, with $J$ the number of $\v y$ variables for each $\v x$ variable. The coupling between the two systems is controlled by the time-independent parameters $h=1$, $c=10$ , and $b=10$. 
Both sets of equations have cyclic boundary conditions, namely $x_{(1)}=x_{(S+1)}$, and $y_{(1)}=y_{(J\cdot S+1)}$.  For most of our experiments the number of state variables are J=32 and S=100  (i.e., the $\v y$ vector has a total of 3200 variables) and to obtain a chaotic behavior, the forcing term $F$ is set to 26. 

The one-scale Lorenz system, 
\begin{equation}
\frac{dx_{(i)}}{dt}=-x_{(i-1)}(x_{(i-2)}-x_{(i-1)})-x_{(i)}+F+ G_{(i)}, 
\label{EQU:Lorenz1S}
\end{equation}
is used as a surrogate model to estimate the true system state from an incomplete set of noisy observations using an ensemble-based data assimilation method. This  introduces model error into our data assimilation and forecasting system since one of the scales is not explicitly represented.

The effect of the missing dynamics (i.e., the effect of fast variables $\v y$) in the surrogate model is approximated by a state dependent parametrization term. As in \cite{pulido16}, $G_{(i)}$ is assumed to be a linear function of the state variable $x_{(i)}$:
\begin{equation}
G_{(i)}=\alpha x_{(i)} + \beta , 
\label{EQU:Lorenz1SPar}
\end{equation}
with $\alpha=19.16$ and $\beta=-0.81$ constant parameters whose optimal values are taken from \cite{scheffler2019}.

The observations were generated from the nature integration every 8 time steps adding a Gaussian error of zero mean and variance equal to $0.2$. Observations are available at  odd grid points (i.e., only 50\% of the system is observed). Given the observations set and the forecasting model, we used the EnKF methodology described in Section \ref{SEC:ENKF} to generate a set of assimilated states $\v {x}^a_k$ that is our best approximation to the real state of the system. 

Two sets of analyses were generated, one using a 100-member ensemble and the other using a 5-member ensemble. In both cases, a localization function was used to reduce the impact of sampling errors. The localization functions follows the one suggested in \cite{GC1999} with a localization scale of 7 grid points which was found to minimize the RMSE of the analyses. These two sets of analyses were used independently to train the neural networks for each experiment as explained in the following sections and as a baseline for analyzing the results. The inflation factor was also tuned to give minimum RMSE, resulting  in an optimal inflation factor of $1.15$ for the 100-member ensemble and $1.35$ for the 5-member ensemble experiment.

The training set consists of 10000 analysis cycles and the validation set has 5000 cycles. The size of the training and validation set is such that converting the Lorenz model time units to atmospheric times is equivalent to 10 years of data. The testing set consists of 15000 time steps which are completely independent from the training and validation sets.

\subsection{NN architecture and training}
\label{SEC:ARCH}

The neural network architecture consists of three convolutional layers (see Table \ref{TABLE:network}). The size of the kernels are relatively small (3 grid points) allowing the identification of patterns in a restricted locality. A kernel width of 5 was also tested but did not result in a better performance that would justify the increase in the network complexity. This is consistent with the behavior of the Lorenz variables that present localized interactions, i.e. two variables that are far from each other have weak interactions. Furthermore, translation invariance is assumed in the convolution layers, which is in accord with the statistical isotropy of the Lorenz'96 dynamics. The size of the output layer depends on the number of subdiagonals of the forecast covariance matrix that are estimated, which in turns depends on the localization matrix $C$ in Eq. \ref{EQU:eMSE_cov}.

\begin{figure}[hbt!]
\centering
        \includegraphics[width=.5\textwidth]{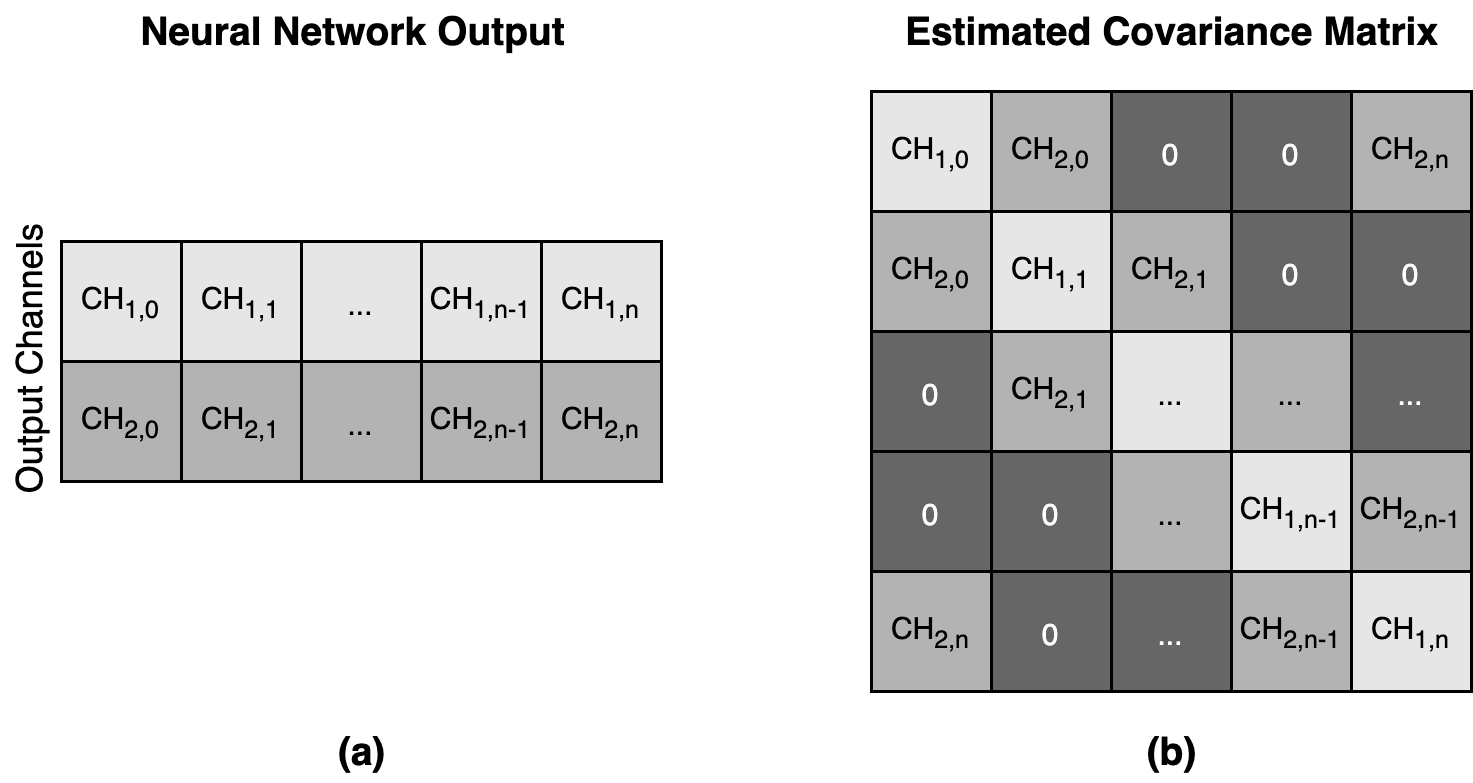}
        \caption{The output of the network is shown in panel \textbf{(a)}, the $i^{th}$ element of channel $0$ is the variance corresponding to the $i^{th}$ state variable. The $d^{th}$ channel corresponds to the covariance between the state variables separated by a distance $d$ as shown in panel \textbf{(b)}. Covariance values not represented with the ANN output are assumed to be zero.} 
\label{FIG:COVLOC}
\end{figure}

For the experiments, we make a simple choice for the localization function $C$. We use a Heaviside function to localize the elements of the target covariance. If the distance between two Lorenz variables is less than $d$ grid points, we leave the corresponding covariance unchanged, and if it is larger than $d$ we set the corresponding covariance to $0$. This is equivalent to keep only the first $n_d$ subdiagonals of the target covariance. In this case we construct our neural-network model to estimate the first $n_d$ subdiagonals of the covariance matrix, while all other subdiagonals are assumed to be $0$ (Figure \ref{FIG:COVLOC}). Sensitivity experiments were carried out to determine the number of subdiagonals needed to optimize the RMSE and to compare this localization value with the optimal localization scale in the EnKF.


\begin{table}
\centering
\begin{tabular}{|c|c|c|c|c|}
\hline
\textbf{layer} &\textbf{in channel} &\textbf{out channel} &\textbf{kernel} &\textbf{activation} \\
\hline
input & 2 & 32 & $3$ & Softplus \\
hidden & 32 & 32& $3$ & Softplus \\
output & 32 & $n_d$ & $3$&Softplus($n_0$)+linear($n_1:n_d$) \\
\hline
\end{tabular}
\caption{Description of the architecture of the convolutional neural network used in the experimentation. The number of channels in the output layer ($n_d$) is the number of subdiagonals to be estimated in the covariance matrix.}
\label{TABLE:network}
\end{table}
As stated in Table \ref{TABLE:network} a single hidden convolutional layer is used. The inclusion of an extra hidden layer did not produce a significant improvement in performance. This convolutional layer use circular padding, since this is consistent with the boundary conditions of the Lorenz model and allows us to keep unchanged the dimensions of the spatial representation through the network. Softplus was chosen as the activation function for the first two convolutional layers since it produces a slightly better convergence among other considered activation functions (viz. logistic and ReLU). In the output layer we use two activation functions: A Softplus  function for the output elements corresponding to the main diagonal ($n_0$) so the estimated variances are positive, and a linear activation function for the elements corresponding to the covariances (subdiagonal elements of the covariance matrix $n_1 \ldots n_d$).  In preliminary experiments, we observed that using linear or ReLU as the activation function in the main diagonal ($n_0$) may lead to the estimation of negative variances or variances equal to zero respectively. 

The AdamW optimizer \citep{adamW} was used to train all the networks with a learning-rate value of $0.001$. The use of mini-batches of $50$ samples produces the best convergence in training. During the training, the loss function is evaluated over the validation set every $10$ training epochs and the training stops when the loss function evaluated over the validation set stops decreasing or starts to increase (early stop with patience). 

Selecting an appropriate network architecture and training hyper-parameters resulted in a challenging task. Network convergence was relatively easy to achieve, however the main difficulty consists on finding a training hyper-parameter set that results in an stable data assimilation cycle using the network estimated covariance. When estimating the forecast covariance, overfitting can be particularly strong since a very noisy target is used and we want the network to filter out the noise and generate a robust estimate of the state-dependent covariance. The risk of overfitting in this setting is much higher than, for example, in the estimation of the mean error. To reduce the risk of  overfitting of the training sample, we use a $L_2$ regularization (weight decay) approach. We noticed that for high values of weight-decay parameter ($\lambda>1$) the variability of the estimated covariance matrix smooths and flattens to an almost state-independent covariance matrix (i.e. the climatological covariance matrix). But once the architecture and the rest of the hyper-parameters have been properly chosen, the best results in terms of performance are achieved with very small values of weight-decay ($1.0e-5$ or even $0$).

\subsection{Forecast error proxies}
\label{SEC:PROXY}

\begin{figure}[hbt!]
\centering
        \includegraphics[width=.5\textwidth]{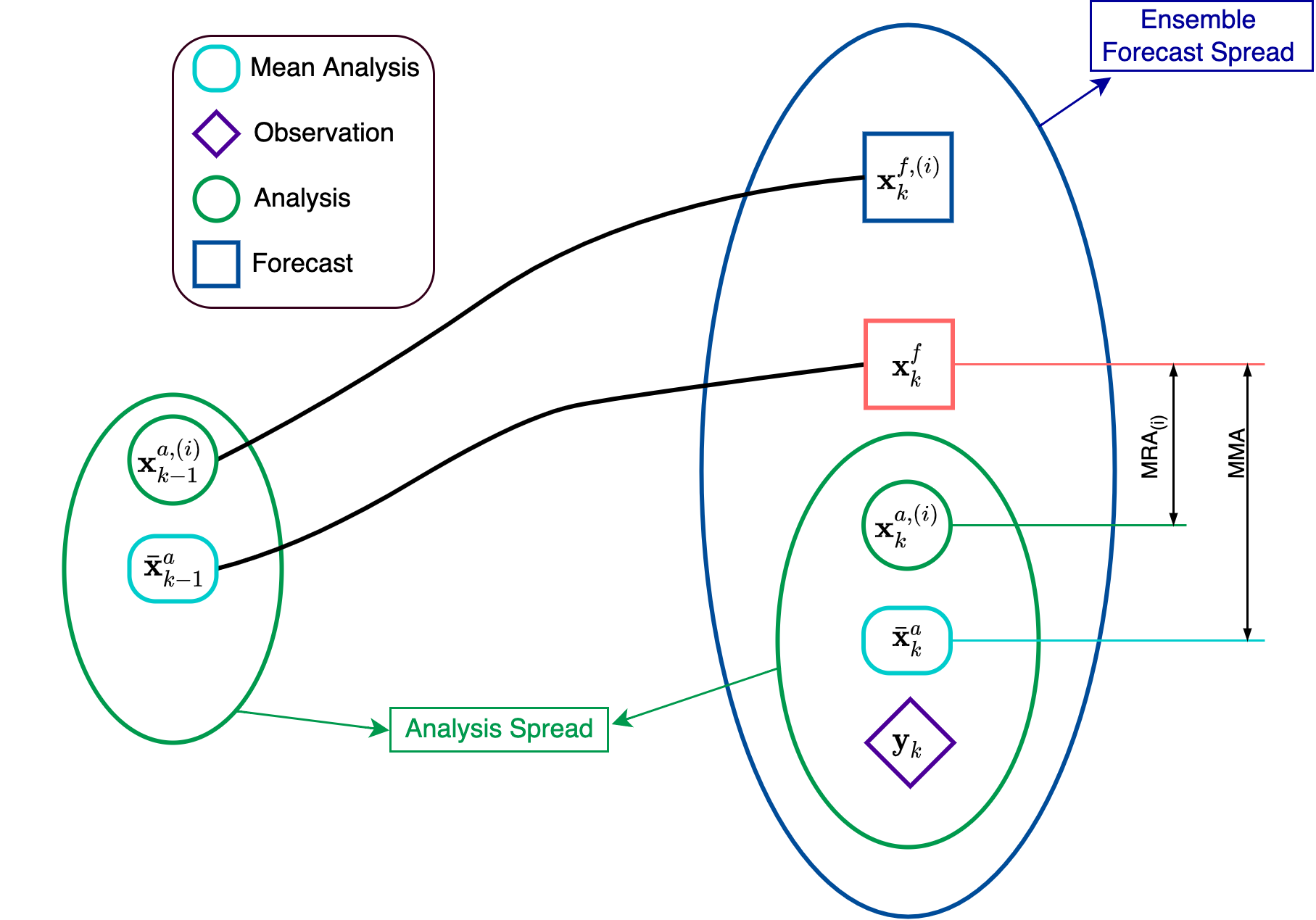}
        \caption{Schematic of an ensemble assimilation cycle where MRA and MMA forecast error proxies are shown using  only the  $i-th$ ensemble member.} 
\label{FIG:DATASET}
\end{figure}


To train the network that estimates the forecast error covariance, we need a dataset that expresses the spatio-temporal variability of the error. But constructing proxies for the short range forecast error (i.e. $\gv\epsilon^f_k$) is a challenging task. In this work, we evaluate different possible proxies for the short range forecast errors, these are schematized in Figure \ref{FIG:DATASET}.
Based on the available ensembles of forecast and analysis, we evaluate three possible ways to define $\gv \epsilon^f_k$:

\begin{itemize}
\item \textbf{M}ean forecast - \textbf{M}ean \textbf{A}nalyisis (\textbf{MMA}): In this case we define $\gv \epsilon^f_k =  \v {x}^f_{k} - \bar{\v x}^a_{k} $ where $\v{x}^f_{k}$ is a deterministic forecast initialized from the analysis ensemble mean ($\v{x}^f_{k} = \mathcal{M}(\bar{\v x}^a_{k-1})$). In this case, we are taking the difference between the most probable state of the system given all the observations up to time $k-1$ and the most probable state of the system given all the observations up to time $k$. 

\item \textbf{M}ean forecast- \textbf{R}andom \textbf{A}nalysis (\textbf{MRA}): The error proxy is defined as $\gv \epsilon^f_k =  \v {x}^f_{k} - \v {x}^{a,(n_r)}_{k}$where $\v {x}^{a,(n_r)}_{k}$ is a randomly selected member from the analysis ensemble. This proxy is assuming that analysis ensemble members represents equally probable realizations of the true state.
It is important to note that once the member is randomly selected, it remains fixed for all training epochs, i.e. the dataset is the same in all epochs and the random selection is done only once. In this formulation it would be possible to augment the training dataset using all the available ensemble members. In fact, enlarging the training set in this way improved the quality of the estimated covariance, and consequently decreased the RMSE of the analyses. However, we chose to use only one randomly selected member for comparison purposes so that the size of the training dataset is the same to the rest of the chosen proxy methods. 



\item \textbf{M}ean forecast- \textbf{N}a\textbf{T}ure (\textbf{MNT}): Since we are conducting idealized experiments, we have access to the true state of the system, thus for evaluating purposes we can compute the true forecast error as $\gv \epsilon^f_k =  \v {x}^{f}_{k} - \v x^{t}_{k}$, where $\v x^{t}_{k}$ is the true system state given by the nature run. This representation of the forecast error cannot be computed in the real applications and is used only for comparison.

\end{itemize}

Other error approximations were evaluated, in particular  $\gv \epsilon^f_k =  \v {x}^{f,(n_r)}_{k} - \bar{\v x}^{a}_{k}$ where $\v {x}^{f,(n_r)}_{k} = \mathcal{M}(\v {x}^{a,(n_r)}_{k-1})$ is a randomly selected member from the forecast ensemble and  $\gv \epsilon^f_k =  \v {x}^{f,(n_r)}_{k} - \v x^{a,(n_r)}_{k}$, i.e. the difference between a randomly selected member of the forecast ensemble and the corresponding member in the analysis ensemble, but none of them gave better results than \textbf{MRA}.

It is important to note that different error proxies are associated with different estimated error variances. Thus, the trace of the estimated covariance matrix using these different proxies to compute the target, can be significantly different. To reduce the impact of this effect in the assimilation cycle and to compare these different approaches in a more consistent way, a multiplicative inflation factor is applied in the data assimilation experiments as in the EnKF. The multiplicative inflation factor is optimized independently for each error proxy using a brute force approach. 

\section{Results}
\label{SEC:RES}

In this section, we present the results obtained with different sensitivity experiments designed to evaluate the performance of the UnnKF and to compare it to the EnKF with two different ensemble sizes, $5$ and $100$ members. Each UnnKF experiment is identified with a name composed of two parts, the first refers to the error proxy used in the training (see section \ref{SEC:PROXY}) and the second one is the number of subdiagonal of the covariance matrix being estimated by the network (including the main diagonal). Data assimilation experiments performed using the EnKF are named as "ENS" followed by the number of members in the ensemble. 


We start by comparing the time evolution of the covariances used in the data assimilation for the $100$-variable Lorenz model with the UnnKF and EnKF methods. The EnKF method uses the sample covariances obtained using $5$-member  (ENS5) and $100$-member (ENS100) ensembles to which a Gaspari-Cohn function with a localisation scale of $7$ grid points has been applied. The UnnKF method uses an ANN  with $n_d=6$ (MRA6) and trained with the MRA error proxy. In all three cases the magnitude of the estimated covariances has been scaled by the optimal multiplicative inflation (i.e., the one that produced the best results in terms of the analysis RMSE).

\begin{figure}[hbt!]
        \includegraphics[width=\textwidth]{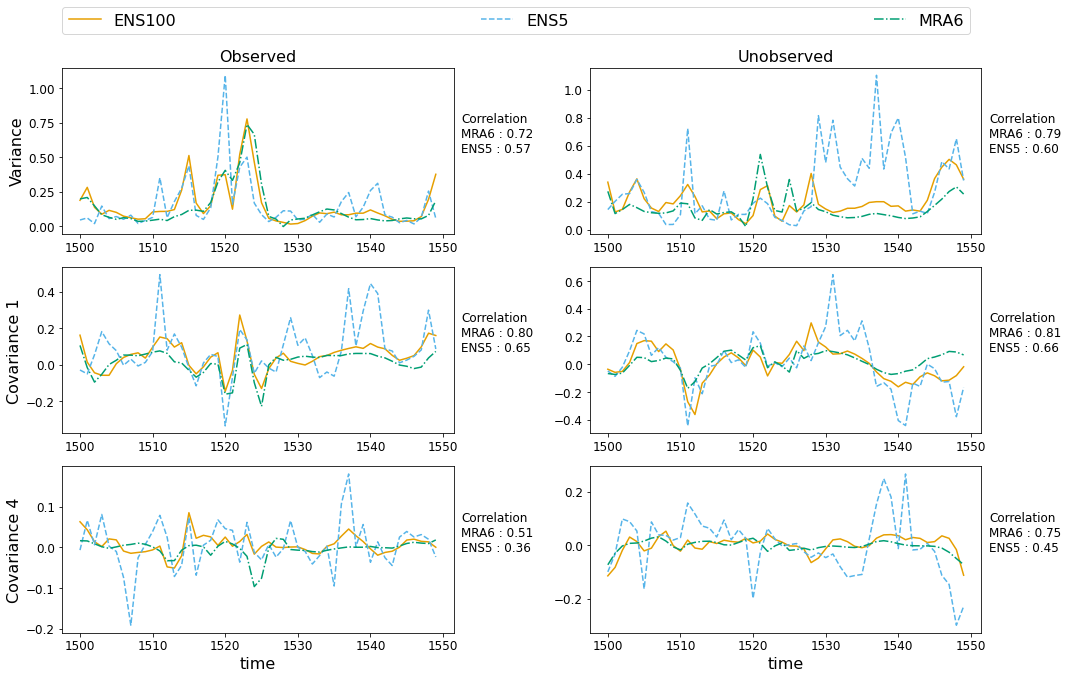}
        \caption{Time evolution of selected covariance matrix elements over 100 consecutive time steps starting at the 1500 cycle of the testing set. The panels, from top to bottom, show respectively the first elements of the first column of the covariance matrix. Left panels show covariance matrix elements of an odd row (centered at observed variables), while right panels shows covariance matrix elements of an even row (centered at an unobserved variable). The correlation coefficient computed over 15.000 assimilation cycles, between ENS5 and MRA6 with respect to ENS100 are shown to the right of each panel. }
\label{FIG:EVOLCOV}
\end{figure}

Figure \ref{FIG:EVOLCOV} shows the time evolution of selected elements of the error covariance matrix as estimated from the EnKF with different ensemble sizes and the neural network. We distinguish between odd covariance matrix rows (centered at an observed variable (Fig. \ref{FIG:EVOLCOV} left column)) and even rows (centered at an an unobserved variable, Fig. \ref{FIG:EVOLCOV} right column)  since their variability can be different.  The temporal correlation coefficient of MRA6 and ENS5 with respect to ENS100 for the entire testing set is stated at the right of each panel of Figure \ref{FIG:EVOLCOV}. 
In all cases, the correlation coefficient of the MRA6 estimate is higher than the correlation of ENS5, even for those estimated covariances which are not shown in the figure. 

The overall analysis shows that ENS5 produces covariances with a higher temporal variability compared to ENS100 due to the effect of sampling noise. In contrast, MRA6 closely follows the variability of the ENS100 for both observed and unobserved variables.
In general, MRA6 has a smoother variability than ENS100 and sometimes it seems to omit some extremes (e.g. time 1515 for the observed variables in all covariances). But it is also able to reproduce quite accurately other extremes present in ENS100 (e.g. variance and covariance at time 1525 for observed variables). Figure \ref{FIG:EVOLCOV} shows that, in general, the time evolution of the covariance matrix estimated by MRA6 is closer to ENS100 than to ENS5. This is consistent with the obtained correlation coefficients already mentioned.  



\begin{figure}[hbt!]
        \includegraphics[width=\textwidth]{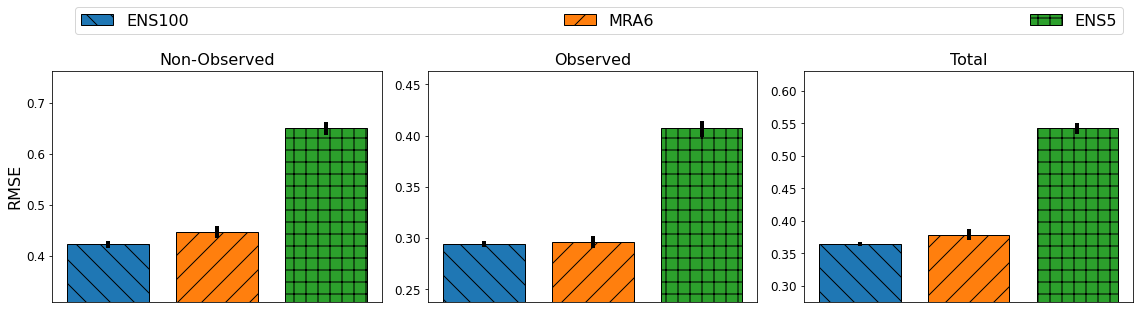}
        \caption{RMSE of the analyses generated by a 5-member ensemble (ENS5), a 100-member ensemble (ENS100) using ENKF method and the UnnKF method (MRA6) for the 100-variable Lorenz'96 model. The RMSE of unobserved variables (left), observed variables (middle) and the total RMSE (right) are shown. Note that the ranges of the RMSE axes are different in each panel, this is to highlight the difference between the experiments. The limits of a $95\%$ confidence interval obtained using a bootstrap approach is indicated by the black line on top of each bar.}
\label{FIG:RMSE_N100}
\end{figure}

To assess the overall quality of the spatio-temporal structure of the estimates in the context of data assimilation, Figure \ref{FIG:RMSE_N100} compares the analysis RMSE over 15,000 consecutive assimilation cycles of the testing dataset using the UnnKF with those generated with ENS5 and ENS100. The black line on top of each bar represent the 95\% confidence interval  computed using a bootstrap approach using  500 subsamples obtained from the testing dataset using random selection with replacement and selecting  samples  which are more than 20 time steps apart from each other to increase the independence between different sample elements.

For both the observed and unobserved variables, the RMSE obtained in Figure \ref{FIG:RMSE_N100} is much closer to ENS100 than to ENS5. This agrees with the time evolution analysis of the covariance matrix elements and shows that the proposed methodology is able to generate a state-dependent estimate of the covariance matrix robust enough to run long assimilation cycles, using only a deterministic forecast as input.


\subsection{Sensitivity to the forecast error proxy}

\begin{figure}[hbt!]
        \includegraphics[width=\textwidth]{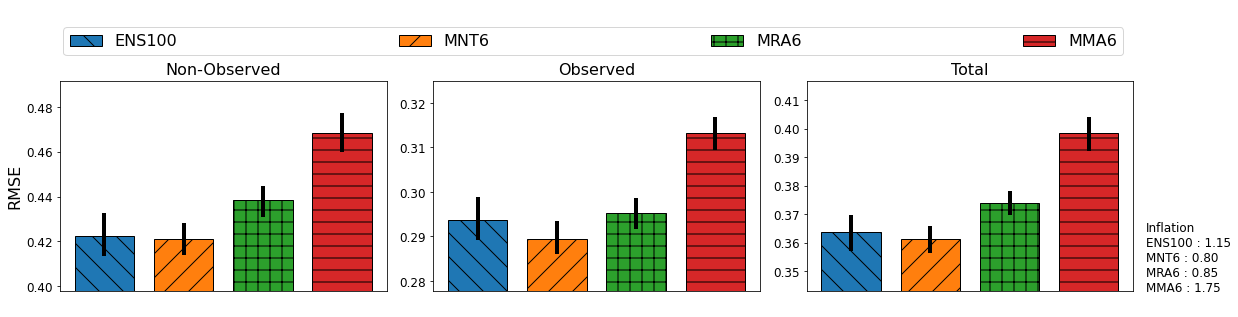}
        \caption{The RMSE of the  unobserved system variables (left panel), the observed variables (middle panel) and the total RMSE (right panel) of the analyses generated using a separate testing dataset are shown for  the networks trained with the datasets,  MNT, MRA and MMA. The limits of a $95\%$ confident interval obtained using a bootstrap approach is indicated by the black line on top of each bar.}
\label{FIG:RMSEDS}
\end{figure}

Figure \ref{FIG:RMSEDS} shows the RMSE of the analysis over the test dataset for an ensamble of 100 member using EnKF and for the UnnKF trained with the actual forecast error (MNT) and the error proxies, MRA and MMA. In all the cases $6$ subdiagonals of the error covariance matrix are estimated. Overall the performance of the UnnKF is stable and produce accurate results. Besides, the skill of the UnnKF is sensitive to the error proxy with roughly a $10\%$ difference between the best (MRA) and the worst (MMA) proxies. 

Using the mean analysis as the training target results in a significantly higher RMSE compared to MRA. Additionally, UnnKF using MMA requires a larger multiplicative inflation factor to achieve optimal performance, indicating that this proxy underestimates the amplitude of errors. This can be explained by the fact that in this case, the forecast error is being approximated as the difference between the deterministic forecast (which is close to the forecast ensemble mean) and the analysis ensemble mean, without considering that the forecast and the real state of the system are random realizations of these distributions. In the MRA approach, a random analysis ensemble member is chosen, taking into account that the analysis members are equally probable realizations of the true state of the system. This proxy overestimates the variance of the errors, leading to optimal multiplicative inflations that are below one.

MNT6 has the lowest analysis RMSE when the estimated covariances are used in the context of a data assimilation system, even better than ENS100. This means that if the error is well estimated, the loss function allows a neural network to be successfully trained to estimate the error covariance. This is quite interesting as it indicates that, given a suitable target, the neural network is able to provide an estimate of the state-dependent uncertainty of the forecast that is more accurate than that provided by a large ensemble, at a much lower computational cost.


The MRA experiment performed similarly to the ENS100 and MNT experiments for the observed variables, but more significant RMSE differences appear for the unobserved variables. Part of these differences can be explained by the effect of model errors. In our experiments, both the forecasts and the analysis are affected by model errors. The assimilation of observations reduces the impact of model errors in the analysis, but some of these errors remain, particularly in the unobserved variables, leading to a misrepresentation of these errors in our proxies.

However, the use of error proxies such as MRA or MMA leads to analysis errors comparable to those obtained with the large ensemble and when the network is trained with the actual forecast error (the analysis RMSE increases by 3.5\% for MRA and 10\% for MMA compared to MNT). This suggests that, despite the limitations of these proxies, they could give reasonable results. The use of MRA, which uses a random member of the analysis ensemble to approximate the forecast error, produces a more accurate representation of the forecast probability distribution than the use of the analysis ensemble mean. The reason for the better performance of the MRA proxy relative to the MMA is not clear. 

The experiment MNT gives an RMSE which is lower than ENS100. The best RMSE for MNT experiments, $0.3593$ (Fig. \ref{FIG:RMSEDS}), is obtained using a multiplicative inflation of 0.85 and a localisation function, as used in the ENS100 experiment, but with a distance of 4 grid points. If no localisation function is used the obtained RMSE for MNT experiment is $0.3655$ and the optimal multiplicative inflation is $0.8$. This indicates a slight overestimation of the distant covariances  in the MNT experiment. On the other hand, the minimum RMSE for the experiments with MMA and MRA error proxies is achieved without applying the localisation function.
\subsection{Sensitivity to localization}
We conducted another  set of experiments to explore the sensitivity of the UnnKF to the number of estimated subdiagonals in the forecast covariance matrix ($n_d$) (i.e., how covariance between distant variables are correctly modeled by the neural network and to what extent the inclusion of covariances between variables that are farther improves the analysis accuracy). The training of these experiments was carried out using the MRA error proxy.

\begin{figure}[hbt!]
        \includegraphics[width=\textwidth]{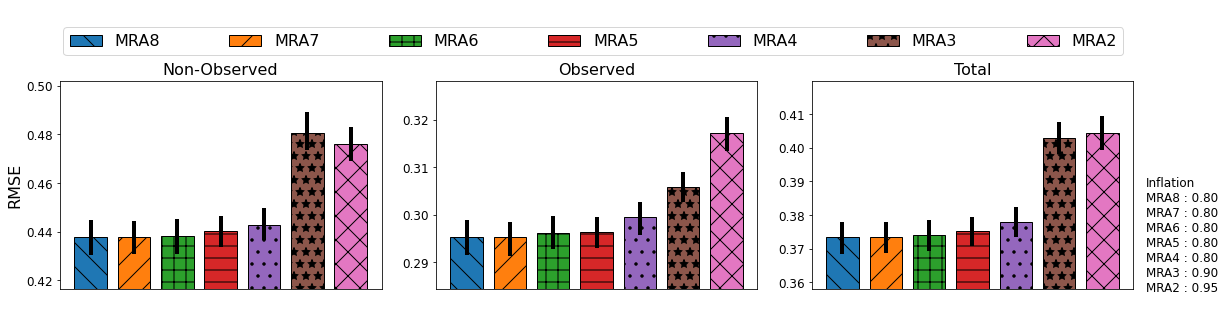}
        \caption{RMSE for band covariance matrices of different sizes $\{2,3,4,5,6,7,8 \}$, for the unobserved system variables (left panel), the observed variables (middle panel) and the total RMSE (right panel) of the analyses generated using a separate testing dataset are shown. The optimal inflation values for each experiments are indicated to the left of the figure. The limits of a $95\%$ confident interval obtained using a bootstrap approach is indicated by the black line on top of each bar.}
\label{FIG:RMSELOC}
\end{figure}

Figure \ref{FIG:RMSELOC} shows the analysis RMSE in the observed, unobserved and total variables as a function of $n_d$. Overall, the larger $n_d$, the lower the RMSE of the analysis with significant reduction of the RMSE up to $n_d=5$. The optimal inflation for each experiment is consistent with the overall performance of the experiment, with larger multiplicative inflation associated with the experiments with larger analysis errors. Beyond $n_d=5$ the RMSE continues to decrease, but the differences are not statistically significant. This shows that the proposed training methodology is able to capture the variability of the farthest covariances containing relevant physical interactions present in the system.   

\begin{figure}[hbt!]
\centering
        \includegraphics[width=.25\textwidth]{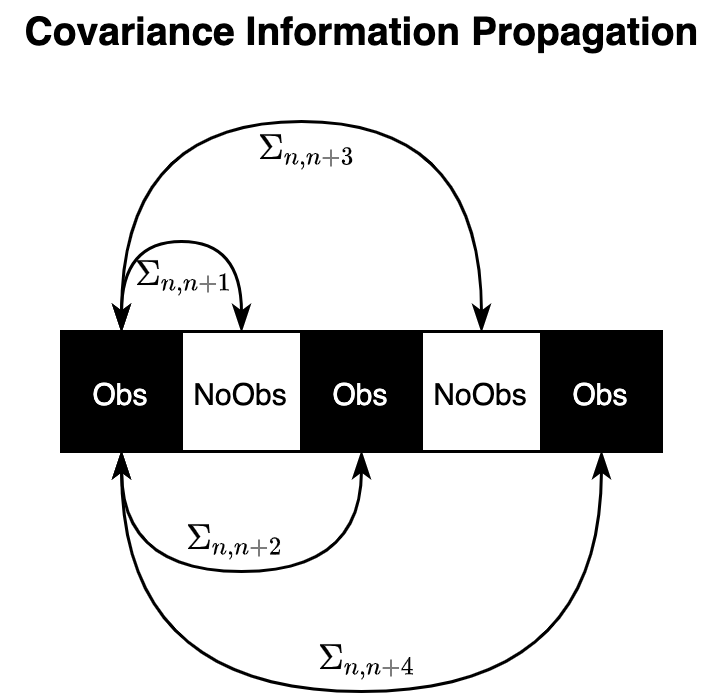}
        \caption{Pattern of observed (\textbf{Obs}) and unobserved (\textbf{NoObs}) state variables during data assimilation process and how each covariance ($\Sigma_{n,i}$) propagates information from one variable ($n$) to another ($i$).} 
\label{FIG:COVPROP}
\end{figure}

For unobserved variables, the experiment with $n_d=3$ does not    lead to an analysis error reduction with respect to the $n_d=2$ case. This can be because, in the experiment with $n_d=3$, the number of observations used to obtain the analysis at unobserved variables is the same as the in the $n_d=2$ experiment. However, the performance on the observed variables improves when $n_d$ is increased from 2 to 3, since the number of observations assimilated at observed variables increases from 1 to 3. This effect is schematized in Figure \ref{FIG:COVPROP} showing that the even subdiagonals propagate information from the observed variables to the unobserved variables, while the odd subdiagonals propagate the information from the observed variables to other observed variables (information is propagated from the observations). This effect seems imperceptible from the 4th subdiagonal onward likely because of the small amplitude of the estimated covariances.

\subsection{Scalability}


In this section we investigate how the optimal size of the neural network (i.e. the number of convolutional filters in the hidden layer) depends on the number of estimated diagonals of the error covariance matrix and on the dimension of the state space. 
We perform experiments varying the size of the neural network and the number of estimated diagonals to evaluate how this affects the RMSE of the analyses.  Results are shown in Table \ref{TABLE:testarch}. It was found that the optimal network capacity is almost insensitive to the number of estimated diagonals ($n_d$). For $n_d=2$ (200 output variables) the optimal number of filters is $32$ and for $n_d=8$ ($800$ output variables) the optimal is $40$. Using larger networks slightly degrades the RMSE. This suggests that adding more channels does not improve  the performance. Furthermore, the increase in output variables related to more subdiagonals in the covariance estimation does not lead to a linear increase in the number of channels. 

\begin{figure}[hbt!]
        \includegraphics[width=\textwidth]{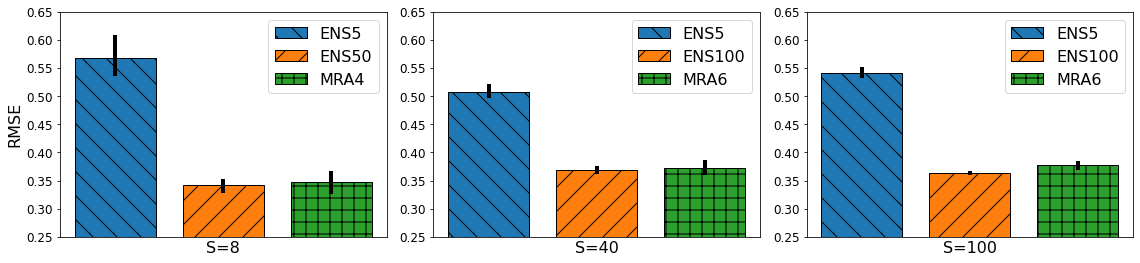}
        \caption{RMSE of the analyses generated for the Lorenz'96 model of 8, 40 and 100 state variables (right, middle and left panels respectively). Each panel shows the RMSE achieved for the EnKF methodology with a small ensemble of 5 members (ENS5), a large ensemble of 50 members for S=8 (ENS50) and 100 members for S=40 and S=100 (ENS100) and for the UnnKF methodology (MRA6).}
\label{FIG:RMSESCAL}
\end{figure}

\begin{table}
\centering
\begin{tabular}{|c|c|c|c|c|}
\hline
\textbf{number of channels} &\multicolumn{2}{c|}{\textbf{$n_d=2$}} &\multicolumn{2}{c|}{\textbf{$n_d=8$}} \\
\textbf{in the hidden layer} & test-loss & RMSE & test-loss & RMSE \\
\hline
10 & 0.11816 & 0.4085 & 0.068481 & 0.39275\\
20 & 0.11758 & 0.4050 & 0.068203 & 0.37823\\
32 & 0.11748 & 0.4042 & 0.068105 & 0.37336\\
40 & 0.11755 & 0.4053 & 0.068130 & 0.37331\\
50 & 0.11766 & 0.4044 & 0.068115 & 0.37335\\
\hline
\end{tabular}
\caption{Loss function and analysis RMSE computed over the testing dataset for different ANN architectures. The first column shows the number of convolutional kernels used in the hidden layer. Results are presented for the case where 2 diagonals ($n_d=2$) and 8 diagonals ($n_d=8$) of the covariance matrix were estimated.}
\label{TABLE:testarch}
\end{table}

We also investigate the sensitivity of the analysis error to the dimension of the state space. 
In this work, we explore the scalability with respect to the state space dimension, by evaluating the performance of the UnnKF for different sizes of the state of the system $S$ (i.e. the number of slow variables in the Lorenz'96 model). We conducted three experiments with $S=8$, $40$ and $100$. For the first experiment (i.e., S=8), we estimated the full covariance matrix ($n_d=4$). For the other two experiments ($S=40$ and $100$), we consider $n_d=6$. These experiments are compared with a small ensemble ($5$-members) and a large ensemble ($50$-members for $S=8$ and $100$-member for $S=40$ and $100$) EnKFs. In all cases, the large ensemble is in the saturation zone of the RMSE curve (i.e., no further significant improvement was obtained by increasing the ensemble size).

The UnnKF has an RMSE that is close to the one of the large ensemble in all the experiments. This indicates that the performance of the estimation of the covariance has no sensitivity to the size of the state vector in these experiments (Fig. \ref{FIG:RMSESCAL}). Moreover, in all cases the performance of the UnnKF is significantly better than the one obtained with the small ensemble size, even though an appropriately tuned covariance localization and multiplicative inflation factor has been used in the EnKF.

\subsection{Sensitivity to target quality}



\begin{figure}[hbt!]
        \includegraphics[width=\textwidth]{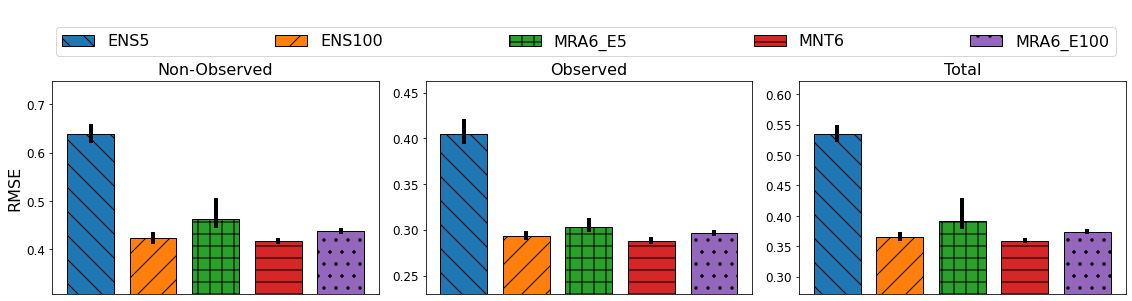}
        \caption{RMSE of different assimilation experiments where ENS100 and ENS5 correspond to the ensemble Kalman filter performance for a 100 and 5 ensemble respectively. MNT6 corresponds to the performance of a neural network trained with the ground truth, MRA6\_E5 is a network trained with the  5-member ensemble dataset and MRA6\_E100 is the network trained with a 100-member ensemble dataset. }
\label{FIG:ESIZEvsRMSEBAR}
\end{figure}

In the experiments presented so far the methodology with the novel loss function achieves a reliable estimate of the state-dependent covariance when trained using an ensemble data assimilation system with a large ensemble. However, in real world applications, available ensemble-based data assimilation systems which can be used for the training of the neural network are based on smaller ensembles due to the high computational cost associated with multiple model integrations. To investigate the impact of the training data quality upon the estimated covariances with the neural network, we performed an additional experiment in which the error proxy is computed from a ensemble-based data assimilation system with only 5 ensemble members. Figure \ref{FIG:ESIZEvsRMSEBAR} compares the results of a neural network in which the error proxy is computed with a 100-member ensemble (MRA6\_E100 as in previous experiments) and the results obtained when the training is performed with an error proxy computed from a smaller 5-member ensemble. The results obtained when the actual forecast error is used for the training are also included for comparison (MNT6).

An  interesting result is that the network trained with error proxies derived from a small ensemble-based data assimilation system  (MRA6\_5) allows us to produce UnnKF analyses with much lower RMSE than the EnKF analyses used in the training data. These results may be explained by the fact that the small ensemble has only a few members to estimate all the elements of the covariance so, sampling errors are expected to be large. However, during training and due to the use of multiple instances at different times the neural network learns to smooth out the different samples leading to a more reliable estimation of the covariance matrix. 

In contrast, analyses generated with large ensembles have very small sampling error but model error is likely to become more dominant. The analyses have a bias, so a part of the forecast error is not well represented during training. Consequently the resulting RMSEs are rather similar with a slight advantage by ENS100 compared to MRA\_100. Meanwhile, the experiment MNT\_100 outperforms ENS100.

\begin{figure}[hbt!]
        \includegraphics[width=\textwidth]{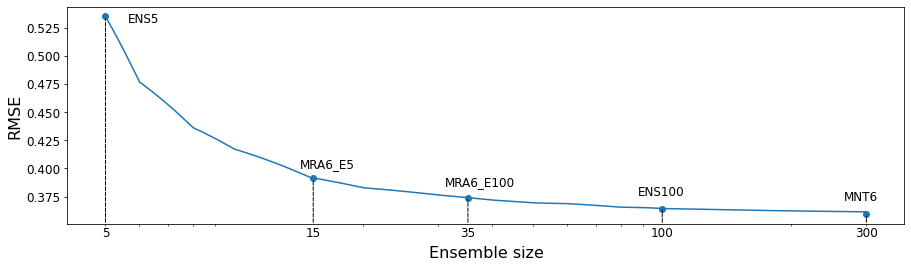}
        \caption{RMSE of the ensemble Kalman filter as a function of the ensemble size. The dots show the RMSE of the different experiments shown in Fig. \ref{FIG:ESIZEvsRMSEBAR} but plotted over the line to relate the performance of the machine learning technique to the ensemble size of the ensemble Kalman filter.}
\label{FIG:ESIZEvsRMSE}
\end{figure}

Figure \ref{FIG:ESIZEvsRMSE} shows the RMSE of the EnKF as a function of the ensemble size. In the same curve, the dots show the corresponding RMSE of the different experiments (i.e., ENS5, ENS100, MRA6\_E100, MRA6\_E5, and MNT6), so that it establishes an equivalence between the accuracy of the machine learning based analysis and the ones produced by the EnKF. In particular, MRA6\_E5 is equivalent to an EnKF with $15$ ensemble-members, MRA6\_E100 is equivalent to a 35-member ensemble, and MNT6 is equivalent to a more than 300 members EnKF. Therefore, we note that MRA6\_E5 improves the performance of the ENS5 dataset. 

\section{Conclusion}
\label{SEC:CONC}
In this work, we examine the potential of machine learning techniques to infer a state-dependent forecast covariance using a single deterministic forecast from a numerical dynamical model. We propose a loss function (eMSE) that allows training a neural network to estimate state-dependent covariance matrices using only previously computed analyses (training target) and current forecasts (model input). Furthermore, this training method is trivially adaptable to localize the covariance matrix to an arbitrary number of diagonals. We also evaluate this novel way to estimate the covariance matrix using a methodology that combines the Kalman-like filter  technique with the neural network covariance estimate (UnnKF), allowing us to perform data assimilation with state-dependent covariance using a single deterministic forecast. Moreover, a model bias correction method could be easily included within the same framework as shown in \cite{Sacco2022}.  
This hybrid data-driven methodology was evaluated in terms of numerical stability and scalability as a function of the size of the state vector. The results are stable (the data assimilation cycle using the UnnKF could be run robustly during $15,000$ cycles) and allowed us to generate analyses with a performance comparable to an ensemble-based data assimilation technique with $100$ members. The optimal network size was not very sensitive to the size of the state space and to the number of covariance matrix elements being estimated, which suggests that the extension of this approach to more realistic applications in high-dimensional state spaces is feasible, although more research is certainly required to confirm this. 

In the experiments where the neural networks was trained with EnKF analysis resulting from a  small ensemble, the UnnKF methodology decreases considerably the RMSE of analyses outperforming the EnKF performance. 
The ability of the UnnKF to outperform its training analysis suggests the possibility to sequentially training it with self-produced analyses generated in a previous training. This process could eventually be repeated until convergence is achieved and the RMSE is stabilised at the lowest possible value. 

Besides the encouraging results there are many challenges and issues that requires further investigation before this methodology can be implemented in combination with state-of-the-art data assimilation systems. For instance, a relatively simple convolutional neural network architecture is used in our experiments. This was sufficient for representing the uncertainty of the two-scale Lorenz-96 dynamics. However, more realistic datasets with multi-scale dynamics are expected to require a deeper network architecture. 
Another important issue is the flexibility of the technique in a context of a continuously changing observing network. In the experiments presented in this work, the observation network is assumed to be fixed. This hypothesis, leads to a quantification of the uncertainty that is implicitly assuming the underlying observation network structure. More work is required to develop more flexible implementations of this methods that can produce a reliable quantification of the forecast error covariance matrix in scenarios where the observing networks are changing. 
More research is also required to more efficiently compute the analysis update. In this work, we conduct an explicit estimation of the analysis update based on the Kalman update equation. However, this approach is not feasible in high dimensions. A local implementation of the UnnKF like the one used in the Optimal Interpolation approach can be used to allow the computation of the analysis in high dimensional systems.

In this work, we analyze a limit case in which only one deterministic forecast run was performed to conduct data assimilation with the UnnKF. However the combination of machine-learning approaches and ensemble-based approaches have been also explored in the literature leading to promising results (e.g. \citealt{gronquist2021}) although the implementation in the context of data assimilation has not yet been tested.  

The results obtained in this proof of concept work, using a simple and numerically stable loss-function, are a first step towards evaluating the potential of hybrid machine-learning data-assimilation techniques that can be applied as operational data assimilation and weather prediction systems in meteorological centers where the computational capacity is limited like in  developing countries where the computational cost of well-established assimilation methods, like 4DVAr or EnKF is prohibitive. Future work will extend and evaluate the present methodology in more realistic datasets.

\bibliography{References}

\begin{thebibliography}{46}
\expandafter\ifx\csname natexlab\endcsname\relax\def\natexlab#1{#1}\fi
\expandafter\ifx\csname url\endcsname\relax
  \def\url#1{\texttt{#1}}\fi
\expandafter\ifx\csname urlprefix\endcsname\relax\def\urlprefix{URL: }\fi

\bibitem[{Amezcua et~al.(2012)Amezcua, Ide, Bishop and Kalnay}]{amezcua2012}
Amezcua, J., Ide, K., Bishop, C. and Kalnay, E. (2012) Ensemble clustering in
  deterministic ensemble kalman filters.
\newblock \textit{Tellus A}, \textbf{64}.

\bibitem[{Bannister(2017)}]{bannister2017}
Bannister, R.~N. (2017) A review of operational methods of variational and
  ensemble-variational data assimilation.
\newblock \textit{Quarterly Journal of the Royal Meteorological Society},
  \textbf{143}, 607--633.
\newblock
  \urlprefix\url{https://rmets.onlinelibrary.wiley.com/doi/abs/10.1002/qj.2982}.

\bibitem[{Bishop(2006)}]{bishop2006}
Bishop, C. (2006) \textit{Pattern Recognition and Machine Learning}.
\newblock Information Science and Statistics. Springer.
\newblock \urlprefix\url{https://books.google.com.ar/books?id=qWPwnQEACAAJ}.

\bibitem[{Bocquet et~al.(2019)Bocquet, Brajard, Carrassi and
  Bertino}]{bocquetetal2019}
Bocquet, M., Brajard, J., Carrassi, A. and Bertino, L. (2019) Data assimilation
  as a learning tool to infer ordinary differential equation representations of
  dynamical models.
\newblock \textit{Nonlinear Processes in Geophysics}, \textbf{26}, 143--162.
\newblock \urlprefix\url{https://npg.copernicus.org/articles/26/143/2019/}.

\bibitem[{Brajard et~al.(2020)Brajard, Carrassi, Bocquet and
  Bertino}]{brajard2020}
Brajard, J., Carrassi, A., Bocquet, M. and Bertino, L. (2020) Combining data
  assimilation and machine learning to emulate a dynamical model from sparse
  and noisy observations: A case study with the lorenz 96 model.
\newblock \textit{Journal of Computational Science}, \textbf{44}, 101171.
\newblock
  \urlprefix\url{https://www.sciencedirect.com/science/article/pii/S1877750320304725}.

\bibitem[{Brajard et~al.(2021)Brajard, Carrassi, Bocquet and
  Bertino}]{brajard2021}
--- (2021) Combining data assimilation and machine learning to infer unresolved
  scale parametrization.
\newblock \textit{Philos Trans A Math Phys Eng Sci}, \textbf{379}, 20200086.

\bibitem[{Buizza et~al.(2022)Buizza, {Quilodr{\'a}n Casas}, Nadler, Mack,
  Marrone, Titus, {Le Cornec}, Heylen, Dur, {Baca Ruiz}, Heaney, {D{\'\i}az
  Lopez}, Kumar and Arcucci}]{buizzaetal2021}
Buizza, C., {Quilodr{\'a}n Casas}, C., Nadler, P., Mack, J., Marrone, S.,
  Titus, Z., {Le Cornec}, C., Heylen, E., Dur, T., {Baca Ruiz}, L., Heaney, C.,
  {D{\'\i}az Lopez}, J.~A., Kumar, K.~S. and Arcucci, R. (2022) Data learning:
  Integrating data assimilation and machine learning.
\newblock \textit{Journal of Computational Science}, \textbf{58}, 101525.
\newblock
  \urlprefix\url{https://www.sciencedirect.com/science/article/pii/S1877750321001861}.

\bibitem[{Burgers et~al.(1998)Burgers, van Leeuwen and Evensen}]{burgers1998}
Burgers, G., van Leeuwen, P.~J. and Evensen, G. (1998) Analysis scheme in the
  ensemble kalman filter.
\newblock \textit{Monthly Weather Review}, \textbf{126}, 1719 -- 1724.
\newblock
  \urlprefix\url{https://journals.ametsoc.org/view/journals/mwre/126/6/1520-0493_1998_126_1719_asitek_2.0.co_2.xml}.

\bibitem[{Camporeale(2018)}]{camporeale1}
Camporeale, E. (2018) Accuracy-reliability cost function for empirical variance
  estimation.

\bibitem[{Camporeale et~al.(2019)Camporeale, Chu, Agapitov and
  Bortnik}]{camporeale2}
Camporeale, E., Chu, X., Agapitov, O.~V. and Bortnik, J. (2019) On the
  generation of probabilistic forecasts from deterministic models.
\newblock \textit{Space Weather}.

\bibitem[{Carrassi et~al.(2018)Carrassi, Bocquet, Bertino and
  Evensen}]{carrassi2018}
Carrassi, A., Bocquet, M., Bertino, L. and Evensen, G. (2018) Data assimilation
  in the geosciences: An overview of methods, issues, and perspectives.
\newblock \textit{WIREs Climate Change}, \textbf{9}, e535.

\bibitem[{Cheng et~al.(2023)Cheng, Quilodran-Casas, Ouala, Farchi, Liu, Tandeo,
  Fablet, Lucor, Iooss, Brajard et~al.}]{cheng2023machine}
Cheng, S., Quilodran-Casas, C., Ouala, S., Farchi, A., Liu, C., Tandeo, P.,
  Fablet, R., Lucor, D., Iooss, B., Brajard, J. et~al. (2023) Machine learning
  with data assimilation and uncertainty quantification for dynamical systems:
  a review.
\newblock \textit{arXiv preprint arXiv:2303.10462}.

\bibitem[{Farchi et~al.(2021{\natexlab{a}})Farchi, Bocquet, Laloyaux, Bonavita
  and Malartic}]{farchietal2021}
Farchi, A., Bocquet, M., Laloyaux, P., Bonavita, M. and Malartic, Q.
  (2021{\natexlab{a}}) A comparison of combined data assimilation and machine
  learning methods for offline and online model error correction.
\newblock \textit{Journal of Computational Science}, \textbf{55}, 101468.
\newblock
  \urlprefix\url{https://www.sciencedirect.com/science/article/pii/S1877750321001435}.

\bibitem[{Farchi et~al.(2022)Farchi, Chrust, Bocquet, Laloyaux and
  Bonavita}]{farchietal2022}
Farchi, A., Chrust, M., Bocquet, M., Laloyaux, P. and Bonavita, M. (2022)
  Online model error correction with neural networks in the incremental 4d-var
  framework.

\bibitem[{Farchi et~al.(2021{\natexlab{b}})Farchi, Laloyaux, Bonavita and
  Bocquet}]{farchietal2021b}
Farchi, A., Laloyaux, P., Bonavita, M. and Bocquet, M. (2021{\natexlab{b}})
  Using machine learning to correct model error in data assimilation and
  forecast applications.
\newblock \textit{Quarterly Journal of the Royal Meteorological Society},
  \textbf{147}.

\bibitem[{Gandin(1965)}]{gandin1965}
Gandin, L. (1965) Objective analysis of meteorological fields:
  Gidrometeorologicheskoe izdatel’stvo (gimiz), leningrad.
\newblock \textit{Translated by Israel Program for Scientific Translations,
  Jerusalem}.

\bibitem[{Gaspari and Cohn(1999)}]{GC1999}
Gaspari, G. and Cohn, S.~E. (1999) Construction of correlation functions in two
  and three dimensions.
\newblock \textit{Quarterly Journal of the Royal Meteorological Society},
  \textbf{125}, 723--757.
\newblock
  \urlprefix\url{https://rmets.onlinelibrary.wiley.com/doi/abs/10.1002/qj.49712555417}.

\bibitem[{Grooms(2021)}]{grooms2021}
Grooms, I. (2021) Analog ensemble data assimilation and a method for
  constructing analogs with variational autoencoders.
\newblock \textit{Quarterly Journal of the Royal Meteorological Society},
  \textbf{147}, 139--149.

\bibitem[{Grönquist et~al.(2019)Grönquist, Ben-Nun, Dryden, Dueben, Lavarini,
  Li and Hoefler}]{gronquist2019}
Grönquist, P., Ben-Nun, T., Dryden, N., Dueben, P., Lavarini, L., Li, S. and
  Hoefler, T. (2019) Predicting weather uncertainty with deep convnets.

\bibitem[{Grönquist et~al.(2021)Grönquist, Yao, Ben-Nun, Dryden, Dueben, Li
  and Hoefler}]{gronquist2021}
Grönquist, P., Yao, C., Ben-Nun, T., Dryden, N., Dueben, P., Li, S. and
  Hoefler, T. (2021) Deep learning for post-processing ensemble weather
  forecasts.
\newblock \textit{Philosophical Transactions of the Royal Society A:
  Mathematical, Physical and Engineering Sciences}, \textbf{379}, 20200092.

\bibitem[{Hamill et~al.(2001)Hamill, Whitaker and Snyder}]{hamilletal2001}
Hamill, T.~M., Whitaker, J.~S. and Snyder, C. (2001) Distance-dependent
  filtering of background error covariance estimates in an ensemble kalman
  filter.
\newblock \textit{Monthly Weather Review}, \textbf{129}, 2776 -- 2790.
\newblock
  \urlprefix\url{https://journals.ametsoc.org/view/journals/mwre/129/11/1520-0493_2001_129_2776_ddfobe_2.0.co_2.xml}.

\bibitem[{H{\"a}rter and {de Campos Velho}(2008)}]{Harteretal2008}
H{\"a}rter, F.~P. and {de Campos Velho}, H.~F. (2008) New approach to applying
  neural network in nonlinear dynamic model.
\newblock \textit{Applied Mathematical Modelling}, \textbf{32}, 2621--2633.
\newblock
  \urlprefix\url{https://www.sciencedirect.com/science/article/pii/S0307904X07002296}.

\bibitem[{Houtekamer and Zhang(2016)}]{houtekamerzhang2016}
Houtekamer, P.~L. and Zhang, F. (2016) Review of the ensemble kalman filter for
  atmospheric data assimilation.
\newblock \textit{Monthly Weather Review}, \textbf{144}, 4489 -- 4532.
\newblock
  \urlprefix\url{https://journals.ametsoc.org/view/journals/mwre/144/12/mwr-d-15-0440.1.xml}.

\bibitem[{Hsieh and Tang(1998)}]{hsiehandtang1998}
Hsieh, W.~W. and Tang, B. (1998) Applying neural network models to prediction
  and data analysis in meteorology and oceanography.
\newblock \textit{Bulletin of the American Meteorological Society},
  \textbf{79}, 1855 -- 1870.
\newblock
  \urlprefix\url{https://journals.ametsoc.org/view/journals/bams/79/9/1520-0477_1998_079_1855_annmtp_2_0_co_2.xml}.

\bibitem[{Hu and Kantor(2015)}]{Humphrey2015}
Hu, H. and Kantor, G. (2015) Parametric covariance prediction for
  heteroscedastic noise.
\newblock In \textit{2015 IEEE/RSJ International Conference on Intelligent
  Robots and Systems (IROS)}, 3052--3057.

\bibitem[{Hunt et~al.(2007)Hunt, Kostelich and Szunyogh}]{HUNT07}
Hunt, B.~R., Kostelich, E.~J. and Szunyogh, I. (2007) Efficient data
  assimilation for spatiotemporal chaos: A local ensemble transform kalman
  filter.
\newblock \textit{Physica D: Nonlinear Phenomena}, \textbf{230}, 112 -- 126.
\newblock Data Assimilation.

\bibitem[{Irrgang et~al.(2020)Irrgang, Saynisch‐Wagner and
  Thomas}]{irrgangetal2020}
Irrgang, C., Saynisch‐Wagner, J. and Thomas, M. (2020) Machine
  learning‐based prediction of spatiotemporal uncertainties in global wind
  velocity reanalyses.
\newblock \textit{Journal of Advances in Modeling Earth Systems}.

\bibitem[{Kalnay(2003)}]{kalnay2003}
Kalnay, E. (2003) \textit{Atmospheric Modeling, Data Assimilation and
  Predictability}.
\newblock Cambridge University Press.

\bibitem[{Kondo and Miyoshi(2019)}]{kondoandmiyoshi2019}
Kondo, K. and Miyoshi, T. (2019) Non-gaussian statistics in global atmospheric
  dynamics: a study with a 10 240-member ensemble kalman filter using an
  intermediate atmospheric general circulation model.
\newblock \textit{Nonlinear Processes in Geophysics}, \textbf{26}, 211--225.
\newblock \urlprefix\url{https://npg.copernicus.org/articles/26/211/2019/}.

\bibitem[{van Leeuwen et~al.(2019)van Leeuwen, Künsch, Nerger, Potthast and
  Reich}]{vanleeuwen2019}
van Leeuwen, P.~J., Künsch, H.~R., Nerger, L., Potthast, R. and Reich, S.
  (2019) Particle filters for high-dimensional geoscience applications: A
  review.
\newblock \textit{Quarterly Journal of the Royal Meteorological Society},
  \textbf{145}, 2335--2365.
\newblock
  \urlprefix\url{https://rmets.onlinelibrary.wiley.com/doi/abs/10.1002/qj.3551}.

\bibitem[{Lguensat et~al.(2017)Lguensat, Tandeo, Ailliot, Pulido and
  Fablet}]{lguensat2017}
Lguensat, R., Tandeo, P., Ailliot, P., Pulido, M. and Fablet, R. (2017) The
  analog data assimilation.
\newblock \textit{Monthly Weather Review}, \textbf{145}, 4093 -- 4107.
\newblock
  \urlprefix\url{https://journals.ametsoc.org/view/journals/mwre/145/10/mwr-d-16-0441.1.xml}.

\bibitem[{Liu et~al.(2018)Liu, Ok, Vega-Brown and Roy}]{Liuetal2018}
Liu, K., Ok, K., Vega-Brown, W. and Roy, N. (2018) Deep inference for
  covariance estimation: Learning gaussian noise models for state estimation.
\newblock In \textit{2018 IEEE International Conference on Robotics and
  Automation (ICRA)}, 1436--1443.

\bibitem[{Lorenz(1995)}]{lorenz96}
Lorenz, E. (1995) Predictability: a problem partly solved.

\bibitem[{Loshchilov and Hutter(2017)}]{adamW}
Loshchilov, I. and Hutter, F. (2017) Decoupled weight decay regularization.
\newblock \urlprefix\url{https://arxiv.org/abs/1711.05101}.

\bibitem[{Ouala et~al.(2018)Ouala, Fablet, Herzet, Chapron, Pascual, Collard
  and Gaultier}]{oualaetal2018}
Ouala, S., Fablet, R., Herzet, C., Chapron, B., Pascual, A., Collard, F. and
  Gaultier, L. (2018) Neural network based kalman filters for the
  spatio-temporal interpolation of satellite-derived sea surface temperature.
\newblock \textit{Remote Sensing}, \textbf{10}.
\newblock \urlprefix\url{https://www.mdpi.com/2072-4292/10/12/1864}.

\bibitem[{Parrish and Derber(1992)}]{parrishandderber92}
Parrish, D.~F. and Derber, J.~C. (1992) The national meteorological center’s
  spectral statistical-interpolation system.

\bibitem[{Pulido et~al.(2016)Pulido, Scheffler, Ruiz, Lucini and
  Tandeo}]{pulido16}
Pulido, M., Scheffler, G., Ruiz, J.~J., Lucini, M.~M. and Tandeo, P. (2016)
  Estimation of the functional form of subgrid-scale parametrizations using
  ensemble-based data assimilation: a simple model experiment.
\newblock \textit{Quarterly Journal of the Royal Meteorological Society},
  \textbf{142}, 2974--2984.

\bibitem[{Rabier et~al.(2000)Rabier, Järvinen, Klinker, Mahfouf and
  Simmons}]{rabier2000}
Rabier, F., Järvinen, H., Klinker, E., Mahfouf, J.-F. and Simmons, A. (2000)
  The ecmwf operational implementation of four-dimensional variational
  assimilation. i: Experimental results with simplified physics.
\newblock \textit{Quarterly Journal of the Royal Meteorological Society},
  \textbf{126}, 1143--1170.
\newblock
  \urlprefix\url{https://rmets.onlinelibrary.wiley.com/doi/abs/10.1002/qj.49712656415}.

\bibitem[{Sacco et~al.(2022)Sacco, Ruiz, Pulido and Tandeo}]{Sacco2022}
Sacco, M.~A., Ruiz, J.~J., Pulido, M. and Tandeo, P. (2022) Evaluation of
  machine learning techniques for forecast uncertainty quantification.
\newblock \textit{Quarterly Journal of the Royal Meteorological Society},
  \textbf{148}, 3470--3490.
\newblock
  \urlprefix\url{https://rmets.onlinelibrary.wiley.com/doi/abs/10.1002/qj.4362}.

\bibitem[{Scheffler et~al.(2019)Scheffler, Ruiz and Pulido}]{scheffler2019}
Scheffler, G., Ruiz, J. and Pulido, M. (2019) Inference of stochastic
  parametrizations for model error treatment using nested ensemble kalman
  filters.
\newblock \textit{Quarterly Journal of the Royal Meteorological Society},
  \textbf{145}, 2028--2045.

\bibitem[{Stanley et~al.(2021)Stanley, Grooms and Kleiber}]{Stanley2021}
Stanley, Z., Grooms, I. and Kleiber, W. (2021) Multivariate localization
  functions for strongly coupled data assimilation in the bivariate lorenz 96
  system.
\newblock \textit{Nonlinear Processes in Geophysics}, \textbf{28}, 565--583.
\newblock \urlprefix\url{https://npg.copernicus.org/articles/28/565/2021/}.

\bibitem[{Tandeo et~al.(2020)Tandeo, Ailliot, Bocquet, Carrassi, Miyoshi,
  Pulido and Zhen}]{tandeoetal2020}
Tandeo, P., Ailliot, P., Bocquet, M., Carrassi, A., Miyoshi, T., Pulido, M. and
  Zhen, Y. (2020) A review of innovation-based methods to jointly estimate
  model and observation error covariance matrices in ensemble data
  assimilation.
\newblock \textit{Monthly Weather Review}, \textbf{148}, 3973 -- 3994.
\newblock
  \urlprefix\url{https://journals.ametsoc.org/view/journals/mwre/148/10/mwrD190240.xml}.

\bibitem[{Tandeo et~al.(2015)Tandeo, Ailliot, Ruiz, Hannart, Chapron, Cuzol,
  Monbet, Easton and Fablet}]{tandeo2015combining}
Tandeo, P., Ailliot, P., Ruiz, J., Hannart, A., Chapron, B., Cuzol, A., Monbet,
  V., Easton, R. and Fablet, R. (2015) Combining analog method and ensemble
  data assimilation: application to the lorenz-63 chaotic system.
\newblock In \textit{Machine Learning and Data Mining Approaches to Climate
  Science: proceedings of the 4th International Workshop on Climate
  Informatics}, 3--12. Springer.

\bibitem[{Terasaki and Miyoshi(2014)}]{Koji2014}
Terasaki, K. and Miyoshi, T. (2014) Data assimilation with error-correlated and
  non-orthogonal observations: Experiments with the lorenz-96 model.
\newblock \textit{SOLA}, \textbf{10}, 210--213.

\bibitem[{Wang et~al.(2018)Wang, Lu, Yan, Luo, Li, Zheng and Zhang}]{wang}
Wang, B., Lu, J., Yan, Z., Luo, H., Li, T., Zheng, Y. and Zhang, G. (2018) Deep
  uncertainty quantification: A machine learning approach for weather
  forecasting.

\bibitem[{Williams(1996)}]{williams96}
Williams, P.~M. (1996) Using neural networks to model conditional multivariate
  densities.
\newblock \textit{Neural Computation}, \textbf{8}, 843--854.

\end{thebibliography}
\end{document}